\documentclass[aps,prx,reprint,preprintnumbers,superscriptaddress,nofootinbib,longbibliography,floatfix]{revtex4-2}
\pdfoutput=1
\usepackage{rotating}
\usepackage{array}
\usepackage{amsmath}
\usepackage[normalem]{ulem}
\usepackage{slashed}
\usepackage{booktabs}
\usepackage[pdftex,table]{xcolor}
\usepackage{units}
\usepackage{xfrac}
\usepackage{mathtools}
\usepackage{empheq}
\usepackage[]{units}
\usepackage{multirow}
\usepackage{amssymb}
\usepackage{url}
\usepackage{comment}
\usepackage{physics}
\usepackage{color,soul}
\usepackage{bbm}
\usepackage[caption=false]{subfig}
\usepackage{adjustbox}
\usepackage[T1]{fontenc}

\usepackage{hyperref}
\hypersetup{
  colorlinks=true,
  citecolor=blue,
  linkcolor=blue,
  urlcolor=blue
}

\newcommand{\omni}{\textsc{OmniLearned}~}
\newcommand{\cosmos}{\textsc{OmniCosmos}~}

\begin{document}

\title{\textsc{OmniCosmos}: Transferring Particle Physics Knowledge Across the Cosmos}

\author{Vinicius Mikuni}
\email{vmikuni@hepl.phys.nagoya-u.ac.jp}
\affiliation{Nagoya University, Kobayashi-Maskawa Institute, Aichi 464-8602, Japan}

\author{Ibrahim Elsharkawy}
\email{ibrahim.elsharkawy@mail.utoronto.ca}
\affiliation{National Energy Research Scientific Computing Center (NERSC), Lawrence Berkeley National Laboratory, Berkeley, CA, USA}
\affiliation{Department of Physics, University of Toronto and Vector Institute, Toronto, ON, Canada}

\author{Benjamin Nachman}
\email{nachman@stanford.edu}
\affiliation{Department of Particle Physics and Astrophysics, Stanford University, Stanford, CA 94305, USA}
\affiliation{Fundamental Physics Directorate, SLAC National Accelerator Laboratory, Menlo Park, CA 94025, USA}

\begin{abstract}
Foundation models build an effective representation of data that can be deployed on diverse downstream tasks. Previous research showed that the \textsc{OmniLearned}, a foundation model for collider physics, could significantly advance discovery potential across collider experiments. In this paper, we show that Foundation Models trained on collider data can transfer to cosmology. Specifically, we improve the prediction of cosmological parameters and halo and galaxy velocities in different datasets from \textsc{CosmoBench}. To the best of our knowledge, this is the first time a collider physics model has been shown to generalize across scientific fields.

%
\end{abstract}

\maketitle

\vspace{10mm}

\section{Introduction}
\label{sec:intro}

Current and near-term cosmological surveys are revolutionizing our understanding of dark matter and dark energy. So far, cosmological insight from survey data has been obtained by first compressing the data and then comparing the low-dimensional summaries to theoretical predictions.  Modern machine learning has catalyzed a movement to go beyond these summaries to perform \textit{field-level analysis} in order to use all of the available information~\cite{leclercq2025fieldlevel}.  While these approaches have the potential to significantly improve the precision of survey data, progress in this direction requires two currently incomplete components: (1) accurate simulations describing all relevant length and time scales and (2) machine learning methods that can accommodate the structure of cosmological data and simulations.  While it may be that machine learning will help address (1), especially beyond the N-body evolution of dark matter, we focus on the second challenge.  In particular, cosmological data are composed of variable-length sets of objects (e.g., galaxies) in space (called a \textit{point cloud} in machine learning), and simulating entire universes is so computationally demanding that inference must proceed with only a relatively small number of synthetic universes. 

Industry-driven neural network architectures have focused on regular grids (for images) and sequences (for text).  One can always voxelize the universe onto a regular grid or sequence to make direct use of such approaches.  Instead, we employ an architecture that is designed for point clouds - it can process variable-length inputs and is invariant under permutations of the inputs (set versus sequence).  State-of-the-art methods of this kind are based on transformers~\cite{DBLP:journals/corr/VaswaniSPUJGKP17}.  While highly effective, transformers also require a lot of data to leverage the capacity of large models.  This is a significant challenge for cosmological inference, where large sets of synthetic universes are unavailable.

We propose to address these challenges by building on a point-cloud foundation model designed for particle collider data.  Like cosmology data, collider data are also variable-length sets. Due to quantum mechanics, each collision produces a stochastic number of outgoing particles, each defined by a three-momentum along with other attributes.  Fully simulating a particle collision is computationally expensive, but much less so than simulating an entire universe.  There are also multiple levels of simulation fidelity.  Our hypothesis is that a sufficiently expressive foundation model for collider physics that is fine-tuned on a relatively small number of cosmological simulations will outperform classical approaches as well as machine learning models trained from scratch on the cosmological simulations.  While the physical laws most relevant for cosmology and those most relevant for particle collisions are not the same, the datasets share enough commonalities to justify this transfer learning objective.

To test this hypothesis, we use the \textsc{OmniLearned} ~\cite{Mikuni:2024qsr,Mikuni:2025tar,Bhimji:2025isp} foundation model and the \textsc{CosmoBench}~\cite{Huang:2025dmm} datasets.  \textsc{OmniLearned} is a point-edge-transformer (PET) trained using supervised and unsupervised learning on one billion labeled and unlabeled hadronic jets from various real and synthetic particle collisions. \textsc{OmniLearned} has been shown to effectively transfer between low and high-fidelity simulations, between simulations and real data, and between different collision systems. In this paper, we demonstrate that the same model can be useful for fundamental physics data that are unrelated to particle jets. The \textsc{CosmoBench} dataset is a set of simulated point clouds for benchmarking tools for inferring parameters in the standard model of cosmology ($\Lambda$CDM). In particular, three different simulators are used to build coarse-grained\footnote{It will be interesting to study how the results change with finer granularity / higher multiplicity cosmological simulations.} universe simulations.  Points represent (effective) dark matter halos or galaxies.  Merger histories are present in one of the datasets, but we focus on the present-time tasks\footnote{It may be interesting to explore connections between pretrained collider models trained using parton shower histories for such tasks in the future.}.  Even though gravity is deterministic, the cosmology simulations are stochastic like the collider simulations because of the initial conditions of the universe.  In addition to data, \textsc{CosmoBench} also includes classical and machine learning benchmark algorithms that we use as references. Most research related to foundation models in particle physics describe the data using specialized tokenization and are trained via self-supervised learning~\cite{Feickert:2021ajf,Birk:2024knn,Hallin:2025ywf,Harris:2024sra,Golling:2024abg,Leigh:2024ked,Bardhan:2025icr}. While these models are also promising candidates for investigating knowledge transfer between particle physics and cosmology, they introduce additional challenges. Because the learned tokenization depends on the set of features used to describe particle interactions, a new tokenizer must be defined for the cosmological data. This contrasts with \textsc{OmniLearned}, where replacing a single layer is sufficient to accommodate the different input representation. Moreover, most tokenizers require an explicit ordering of the inputs. Since the ordering of the input halos should not affect the regression performance, a point-cloud representation that is permutation invariant by design is better aligned with the structure of the data.

A number of related works have touched upon some aspects relates to our research.  For example, a number of studies have explored point cloud / graph-based machine learning models applied to cosmology~\cite{ravanbakhsh2016estimating, makinen2022cosmic, villanueva2022learning, desanti2023robust, roncoli2023domain, massara2023predicting, shao2022robust, cuesta2024point, jespersen2022mangrove, wu2023learning, chuang2024leaving, wu2024galaxy, nguyen2024dreams, villanueva2022inferring, cranmer2021unsupervised, wang2022graph, park2023hierarchical, chatterjee2025cosmology, balla2024cosmic,Krishnaraj:2025fye}. \textsc{CosmoBench} includes one such model, and we also compare our foundation model to a setup with the same architecture, but trained from scratch.  This allows us to disentangle benefits from an optimized architecture and from pre-training.  The philosophy of \textsc{OmniLearned} is to encode most of the physics in the model implicitly through pre-training.  Explicitly equivariant models have been proposed for cosmological inference~\cite{perraudin2019deepsphere,defferrard2019deepsphere,cohen2018spherical,ocampo2023scalable,esteves2020spin,krachmalnicoff2019cmb,petroff2025bayesian,dai2022trenf,jagvaral2025geometric,cuesta2025scorematching,thiele2022tsz,minartz2024equivariant,villar2021scalars,casas2025darkmatter,weiler2018steerable3d,gerken2023geometric} and it would be interesting to explore in the future if pre-training outperforms them as it seems to for colliders~\cite{Bhimji:2025isp}.  There are many active efforts to create foundation models specific to cosmology~\cite{xia2025mosaic, parker2024astroclip, parker2025aion, leung2024foundation, nguyen2023astrollama, walmsley2022galaxy, walmsley2024scaling}, but we are not aware of any trained to ingest entire universes-as-point-clouds.  The approach to uncertainty quantification in \textsc{CosmoBench} is relatively simple and it would be interesting to explore the interplay between advanced UQ (see eg Ref.~\cite{Chakkappai:2025noy,fair_universe_wl_challenge_2025} and references within) and foundation models in the future.

The paper is organized as follows. Sec.~\ref{sec:pet} describes the \omni model architecture and the modifications used to adapt to cosmology problems.
Sec.~\ref{sec:dataset} then introduces the dataset used to train \cosmos. Results for different tasks and datasets are then described in Sec.~\ref{sec:results}.  The paper ends with conclusions and outlook in Sec.~\ref{sec:conclusions}.

\section{Knowledge Transfer}
\label{sec:pet}

In particle physics, the set of reconstructed particles produced after a collision is used to study how fundamental particles interact with each other. At high-energy colliders like the LHC, a ubiquitous object produced through particle interactions is a jet, or a collimated spray of particles. Produced through the interactions of colored particles via the strong force, jets are fundamental research objects as they provide context regarding the properties of Quantum Chromodynamics (QCD) as well as the underlying physics process that emerges from the particle collision. Due to their abundance and complexity, many machine learning applications in particle physics were initially proposed for jets~\cite{Larkoski:2017jix,Kasieczka:2019dbj,Feickert:2021ajf}, including CNNs~\cite{deOliveira:2015xxd}, RNNs~\cite{Guest:2016iqz}, GNNs~\cite{Henrion:DLPS2017,Qu:2019gqs}, GANs~\cite{deOliveira:2017pjk}, Deep Sets~\cite{Komiske:2018cqr}, and attention~\cite{Mikuni:2020wpr}/transformers~\cite{Mikuni:2021pou,Qu:2022mxj}.  The current state of the art~\cite{ Qu:2022mxj,Wu:2024thh,Brehmer:2024yqw,Bhimji:2025isp} represents jets as point clouds, or unordered sets of objects in a metric space.  These point clouds are processed with transformer models which can account for long range correlations between particles and can be scaled effectively to bigger models trained on more data.  
%

\begin{figure*}[ht]
    \centering
        \includegraphics[width=.9\textwidth]{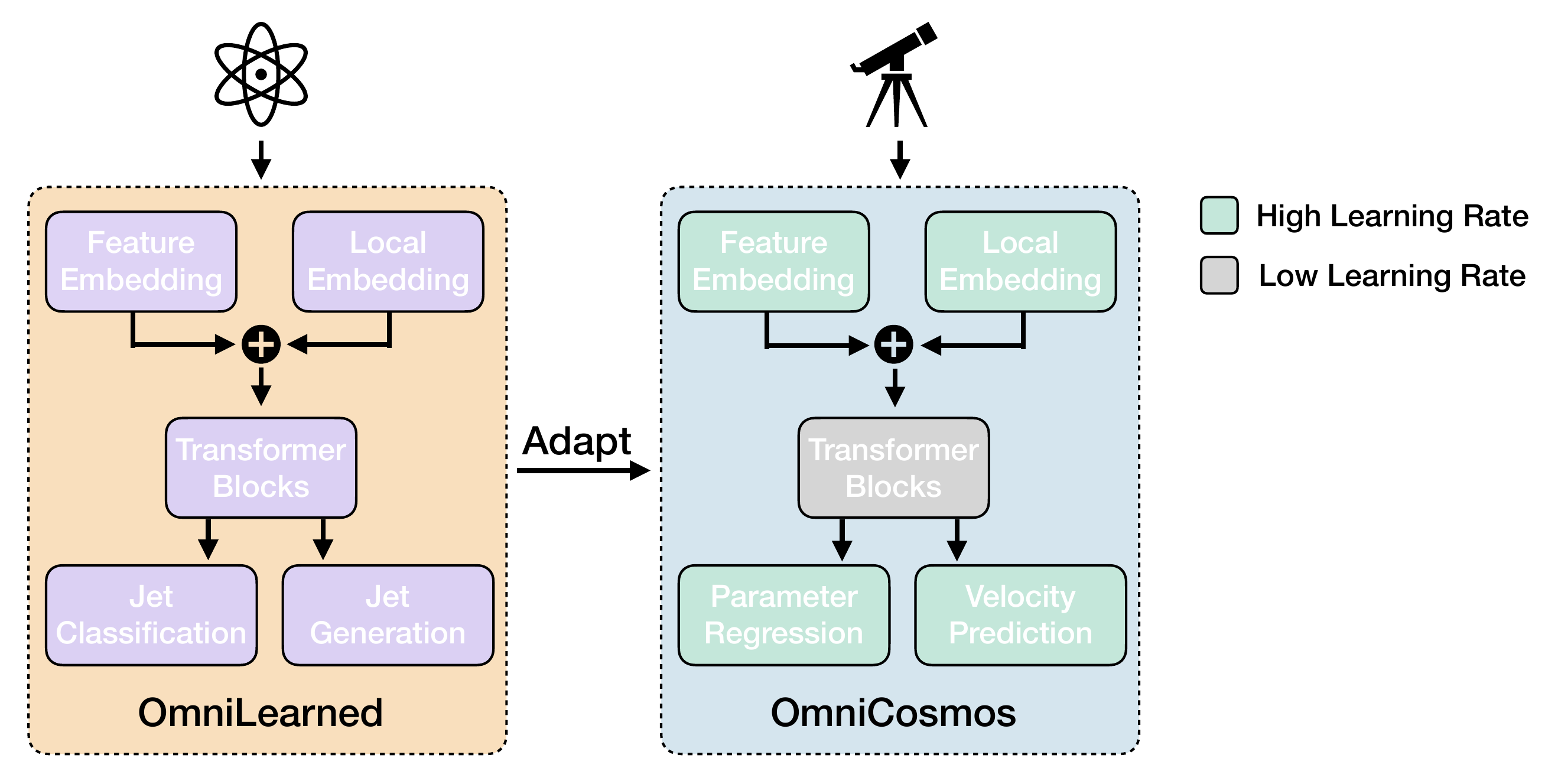}
    \caption{Summary of the knowledge adaptation strategy between the \textsc{OmniLearned} model, trained on particle collisions, to cosmology tasks named \textsc{OmniCosmos}. }
    \label{fig:omnicosmos}
\end{figure*}

The \omni model~\cite{Mikuni:2024qsr,Mikuni:2025tar,Bhimji:2025isp} uses particles clustered into jets as part of the inputs and employs different tasks to create a rich point cloud representation that can be efficiently adapted to different datasets and tasks. The underlying neural network model, named \textsc{PET-v2}, consists of a mixture of local and global transformer layers tasked to learn local and global correlations between particles. Crucially, generic transformer layer logits $A_{ij}$ are augmented using physics-inspired interaction bias terms $B_{ij}$, 
\begin{equation}
A_{ij}^{*} \;=\; A_{ij} \;+\, B_{ij}, \quad
B_{ij} \in \mathbb{R} \;\; 
\label{eq:attn-bias}
\end{equation}
shown to improve performance even in the presence of large training datasets. The model is then trained using 1 billion jets containing a mixture of real particle collisions, as well as simulated particle interactions with different degrees of simulation fidelity. The result of this pre-training stage represents the core foundation model that we refer to \omni. To adapt to specific applications, we fine-tune \omni to the specific dataset and task to be solved. The first challenge is to identify an efficient strategy to adapt \omni to a completely different dataset. Whereas particles used as inputs to \omni are described by their four-momenta, the dark matter halo simulations used in this work are described by their spatial location in 3D coordinates. Moreover, the relevant tasks considered in the pre-training phase of \omni are point cloud classification and generation, while in this case, we are interested in the regression of different cosmological parameters and the halo velocity prediction for each input object. Last, an average jet contains $\mathcal{O}(10-100)$ particles while we consider up to 5000 of the most massive halos in the \textsc{CosmoBench} datasets. We address all these challenges by introducing \cosmos, with a general adaptation strategy shown in Fig.~\ref{fig:omnicosmos}

First, we address the differences in object representation between point clouds. In the \textsc{PET-v2} model, the initial input features for each point in the point cloud are transformed using a feature embedding that maps the inputs to a large latent space. Fully connected layers with non-linear activations (FCs) take as inputs only the initial feature representation, applying the same transformation to all points in the point cloud to ensure permutation equivariance of the outputs. 

A local embedding that includes the local particle information is determined in parallel using  the information of the k-nearest neighbors of each input point. For each object $\vec x_i$, we identify the k-nearest neighbors $\vec x_j$ using the Euclidean distances in input space. Next, we augment the original input features by including relevant functions $f(\vec x_i,\vec x_j)$ calculated between pairs of points. At this step, we are able to include tailored functions specific to the physics problem and dataset used. While in \omni we include several features that are relevant for particle physics, such as the Lorentz invariant mass of the combination of two particles, we instead identify three geometrical features to be included in addition to the difference $\vec x_i - \vec x_j$.  We include the absolute distance $d(\vec x_i, \vec x_j) = \lVert \vec x_i - \vec x_j \rVert_2$, the cosine distance $d_{cos}(\vec x_i, \vec x_j) = 1 - \frac{\vec x_i \cdot \vec x_j}{\lVert \vec x_i \rVert_2 \, \lVert \vec x_j \rVert_2}$, and the cosine distance between the initial halo $\vec x_i$ and the difference $\vec x_i - \vec x_j$, described by $d_{cos}(\vec x_i, \vec x_i - \vec x_j) = 1 - \frac{\vec x_i \cdot (\vec x_i -\vec x_j)}{\lVert \vec x_i \rVert_2 \, \lVert \vec x_i - \vec x_j \rVert_2}$. The resulting pairwise features considered are:

\begin{equation}
\begin{split}
    f(\vec x_i,\vec x_j) = [\vec x_i - \vec x_j, d(\vec x_i, \vec x_j),d_{cos}(\vec x_i, \vec x_j), \\d_{cos}(\vec x_i, \vec x_i - \vec x_j)].
\end{split}
\end{equation}

These pairwise features are used as inputs to a transformer block, where the attention mechanism attends to only particles $\vec x_i$ and their $k$-neighbors. Last, we take the average response across the k-neighbors to define the new representation for each point. 

The final latent representation for each point in the point cloud is the sum of the two previously described operations, the one that applies the same learnable function to all particles, and the one that leverages the local neighborhood.

To create \cosmos with \omni pre-training, we start by loading all the pretrained weights of \omni in the new network architecture. Whenever weights are incompatible due to differences in input shapes, randomly initialized weights are used. These correspond to only 0.09\% of the total model weights. Since the new layers require more training steps for convergence, we set the learning rate of the newly initialized layers higher than the learning rates of pre-trained layers. This also has the effect of properly retaining the strong geometric priors learned in pre-training while preventing "catastrophic forgetting."  
 
The second challenge is to address the new training tasks. Whereas \omni focuses on classification and particle generation, we are interested in regressing cosmological parameters and predicting the three-dimensional halo velocity for each object in the point cloud. While these tasks are remarkably different in scope, they are similar in output. In the classification task, we predict class scores for each jet based on all available classes. For regression, the only modification needed is to change the output size to match the number of parameters to predict. For the velocity prediction, we can re-purpose the particle generator network, which originally assigned an output prediction per particle, to instead predict the halo velocities. In this case, again, we only need to replace the last layer with a new one to correct for the difference in output shape. This layer is newly initialized, and thus the learning rate assigned is also higher. 

The last challenge is to address the difference in the number of points between the original jet data and the current datasets. Since \omni does not enforce a fixed particle multiplicity, no additional modification is required. However, the tenfold increase in the number of particles restricts the number of transformer blocks that can be used due to memory requirements. Because of this, we study only the adaptation of the small \omni model, with eight transformer blocks and a total of around two million trainable parameters. 

The result of the knowledge adaptation between the \omni model, initially trained on particle collisions, is the \cosmos model, used to address multiple tasks in cosmology.

\section{\textsc{CosmoBench} Dataset}
\label{sec:dataset}

The \textsc{CosmoBench}~\cite{Huang:2025dmm} dataset introduces large-scale cosmological simulations of dark matter halos and galaxies using multiple physics simulators, amounting to more than 41 million core-hours to produce. In this study, we consider two datasets from \textsc{CosmoBench}: the \textsc{Quijote}~\cite{Villaescusa-Navarro:2019bje} and \textsc{CAMELS-SAM}~\cite{Perez:2022nlv} simulation suits. In both datasets, we are interested in estimating the values of cosmological parameters $\Omega_m$ and $\sigma_8$, representing the fraction of the total energy density of the universe made up of matter (including both dark matter and normal matter) and the amplitude of matter density fluctuations, respectively. From random initial conditions, the different programs use N-body simulations to evolve the systems to present time. The \textsc{CAMELS-SAM} simulations consist of dark-matter-only N-body
simulations post-processed with a semi-analytic galaxy-formation model with a box size of 100 cMpc$ / h$ and dark matter particle mass of $\sim 10^8$M$\odot / h$. The \textsc{Quijote} simulations use a box of size 1000 cMpc$ / h$ and consider dark matter particles of masses $\sim 10^{12}$M$\odot / h$. Due to the lower resolution, the \textsc{Quijote} simulations are cheaper to create compared to \textsc{CAMELS-SAM}, where the higher mass resolution allows it to resolve smaller, highly non-linear scales, at the expense of simulating a smaller volume. In both datasets we use up to 5000 most massive halos available in the simulations. We use the official dataset split for each dataset, consisting of 600 / 204 / 196 samples for training, validation, and testing in the \textsc{CAMELS-SAM} dataset and 19651 / 6550 / 6551 for the \textsc{Quijote} simulations. 


\section{Results}
\label{sec:results}

\begin{figure*}[ht]
    \centering
        \includegraphics[width=.47\textwidth]{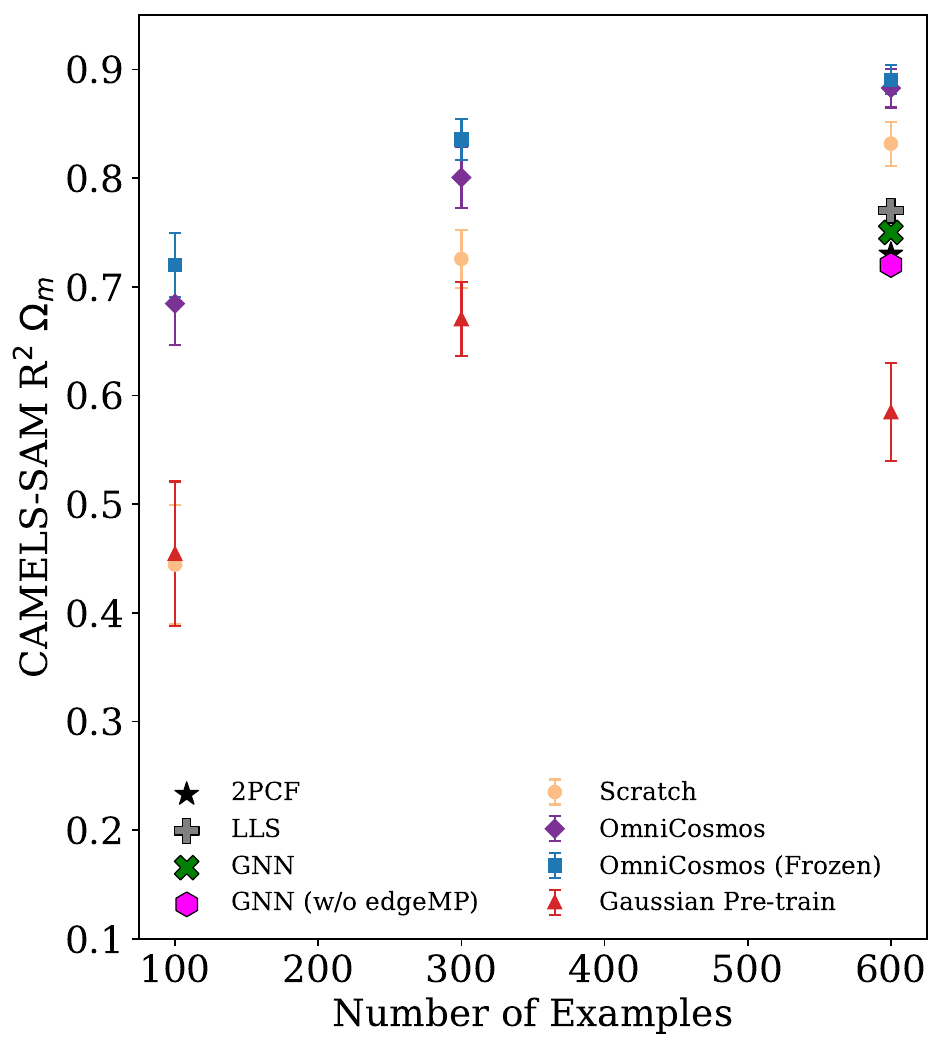}
        \includegraphics[width=.47\textwidth]{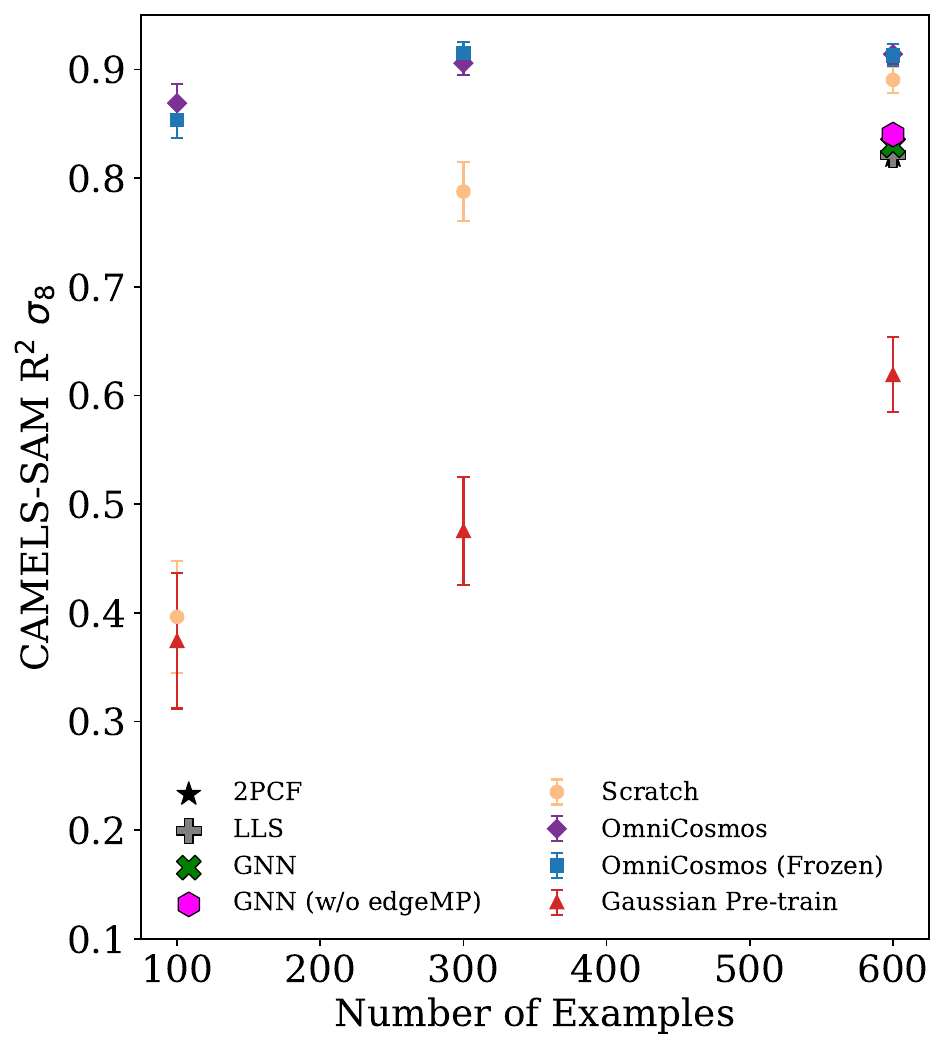}
    \caption{Results for the prediction of different cosmological parameters obtained using the \textsc{CAMELS-SAM} simulations.}
    \label{fig:camels_parameter}
\end{figure*}

We evaluate the performance of \cosmos to regress cosmological parameters in the \textsc{CAMELS-SAM} simulations using different amounts of available training samples. We quantify the performance using the coefficient of determination $R_y^2$ for a parameter $y$ determined as:

\begin{equation}
    R_y^2 = 1 - \frac{\sum_{i=1}^n (\hat{y}_i - y_i)^2}{\sum_{i=1}^n (\bar{y_i} - y_i)},
\end{equation}

where the model prediction $\hat{y}$ is compared against the true value $y$ with mean value $\bar{y}$ for each entry $i$ in the test dataset. We determine the uncertainties on $R_y^2$ by bootstrapping the test data 1000 times with replacement, reporting the mean value as the final prediction and the standard deviation of the bootstraps as the uncertainty. We directly compare our results with the ones provided in~\cite{Huang:2025dmm} without retraining. 

Inputs are standardized before the training. We test different hyperparameters for both \cosmos and the model trained from scratch, i.e, with all trainable weights initialized from random weights.  Since the network parameters are fixed from the original \omni model, we consider different values of learning rate, batch size, and the multiplicative factor between the lower learning rate and higher learning rate layers in the \cosmos fine-tuning. The settings corresponding to the best-performing models are reported in the final results. To determine the size of the local neighborhood from each point, we use the value of $k = 10$. The results for the prediction of both $\Omega_m$ and $\sigma_8$ are shown in Table~\ref{tab:camels_parameter} and Fig.~\ref{fig:camels_parameter} as a function of the number of examples in the training set. In addition to the comparison between the model trained from scratch and \cosmos we consider two additional ablation results. One where the transformer layers of the original \omni model are frozen, i.e are not updated during the adaptation step, and one where the same architecture is first pre-trained on a synthetic point cloud. The first ablation is used to study how useful are the learned representations from the \omni model when adapting to new datasets, the second is to understand how useful is the particle physics representation used to pre-train the model. The synthetic point cloud consists of ten equally populated classes, with each class represented by four independent Gaussian-distributed features. For class $c \in \{0,\ldots,9\}$ and feature $f \in \{0,\ldots,3\}$, the corresponding Gaussian mean and standard deviation are defined as
\begin{equation}
\mu_{c,f} = -2.5 + \frac{4c}{9} + \frac{f}{3},
\end{equation}
and
\begin{equation}
\sigma_{c,f} = \left( 0.5+\frac{c}{9} \right) \left( 0.8+\frac{0.4f}{3} \right).
\end{equation}

Both the feature means and standard deviations increase monotonically with the class label and result in features with comparable ranges as the ones used to pre-train the \omni model. As in the original \omni model, pre-training jointly targets two tasks: classifying the input point clouds and generating them. The synthetic pre-training dataset contains 10 million point clouds, each composed of 150 points, chosen to match the typical point cloud size used to pre-training the \omni model. During the adaptation phase, the transformer blocks of this pre-trained model are also kept frozen to enable the direct comparison with the \omni model.

\begin{table}[th]
    \centering
    \caption{Comparison between the performance reported for different algorithms on the \textsc{CAMELS-SAM} dataset. Bold results represent the algorithm with highest performance.}
    \label{tab:camels_parameter}
	\begin{tabular}{lccccc}
    \hline
          $\text{R}^2 \uparrow$&  $\Omega_m$ & $\sigma_8$ \\
            \hline
            2PCF & 0.73 $\pm 0.03$ & 0.82 $\pm 0.02$\\
            LLS & 0.77 $\pm 0.03$ & 0.82 $\pm 0.02$ \\
            GNN & 0.75 $\pm 0.03$ & 0.83$\pm 0.02$ \\
            GNN (w/o edgeMP) & 0.72 $\pm 0.03$ & 0.84$\pm 0.02$ \\
            \hline
            Scratch & 0.83 $\pm 0.02$ & 0.89 $\pm 0.01$   \\
            \textsc{OmniCosmos} & \textbf{0.88 $\pm$ 0.01} & \textbf{0.91 $\pm$ 0.01}   \\
            \textsc{OmniCosmos} (Frozen) & \textbf{0.89 $\pm$ 0.01} & \textbf{0.91 $\pm$ 0.01}   \\
            Gaussian Pre-train & 0.58 $\pm$ 0.04 & 0.61 $\pm$ 0.03   \\
	
	\end{tabular}
\end{table}

\begin{figure}[ht]
    \centering
        \includegraphics[width=.45\textwidth]{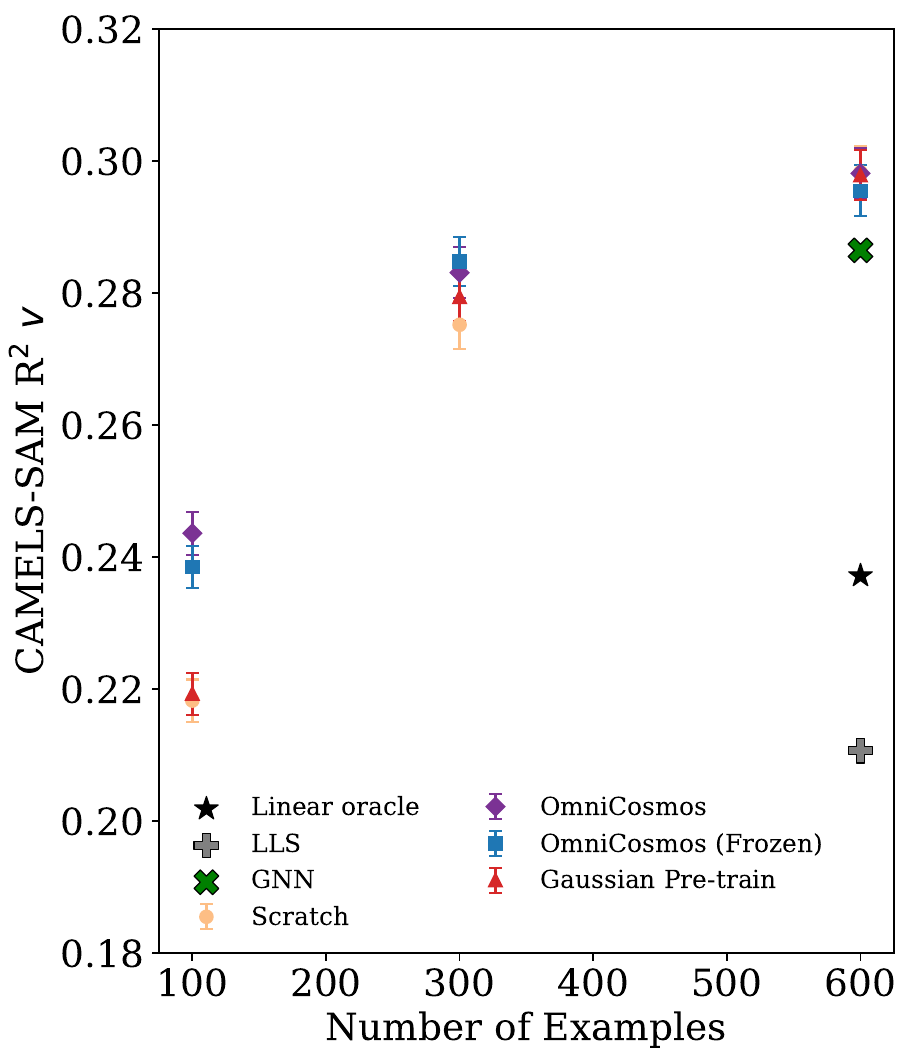}
    \caption{Results for the halo velocity prediction using the \textsc{CAMELS-SAM} simulations.}
    \label{fig:camels_velocity}
\end{figure}

\begin{figure*}[ht]
    \centering
        \includegraphics[width=.47\textwidth]{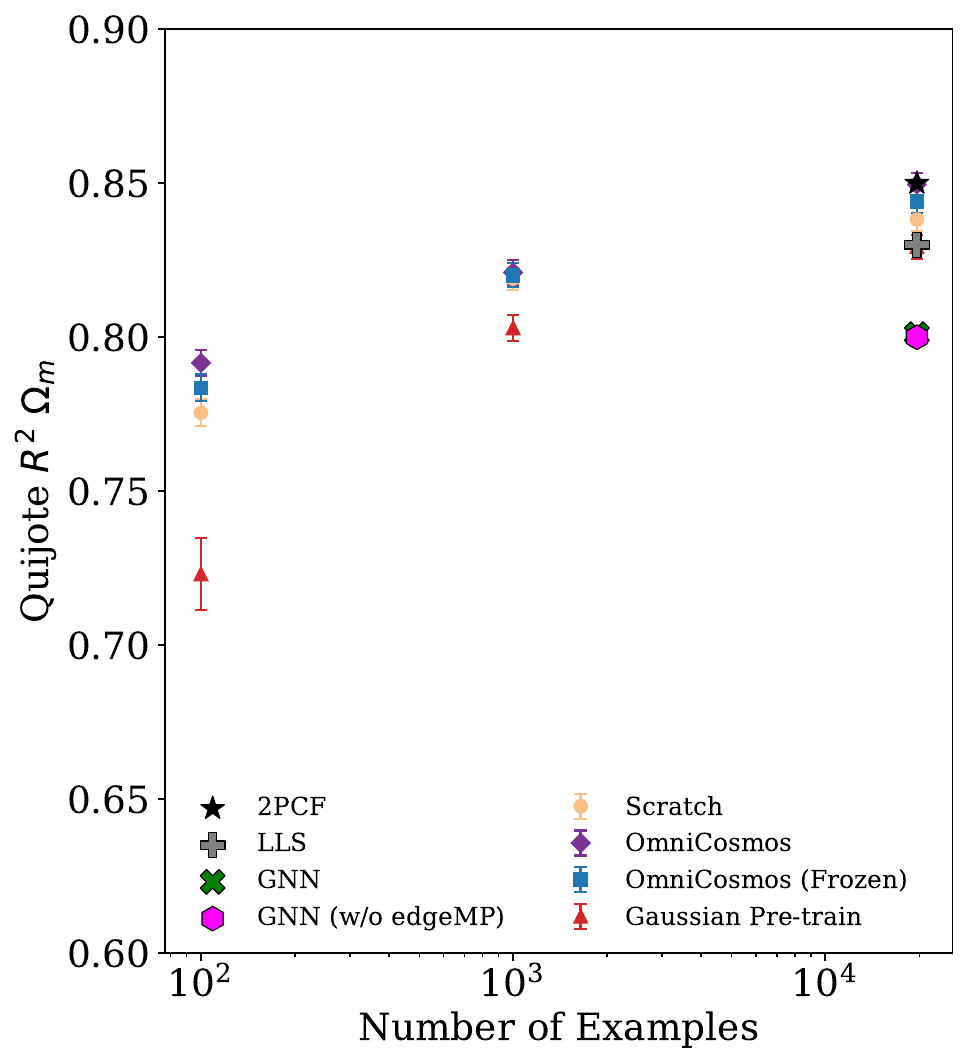}
        \includegraphics[width=.47\textwidth]{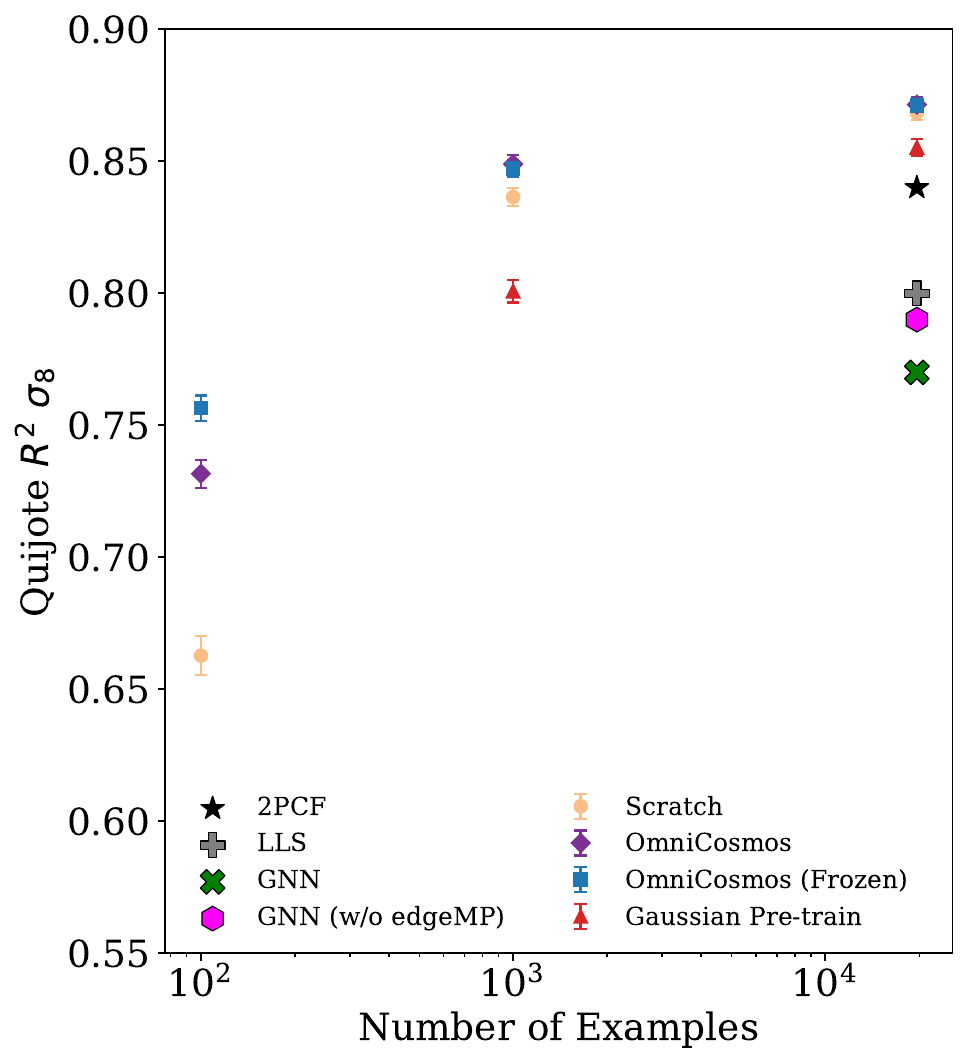}
    \caption{Results for the prediction of different cosmological parameters obtained using the \textsc{QUIJOTE} simulations.}
    \label{fig:quijote_parameter}
\end{figure*}

We observe the performance obtained by both the fine-tuned \cosmos model and the model trained from scratch to surpass previous benchmarks when using the full available training dataset. Interestingly, with only 100 examples of the available simulations \cosmos is already able to match or improve upon the performance from previous benchmarks. Moreover, we see a strong benefit from the fine-tuning strategy as opposed to training from scratch, where at all dataset sizes, the \cosmos model is able to obtain better predictions of the parameters, showing promising results even at the very small dataset size of 100 simulations. We also observe that the full fine-tuning strategy and the strategy in which the transformer blocks are frozen yield similar results within uncertainties. This indicates that the representations learned during pre-training of the \omni model are also useful for cosmology. This conclusion is further supported by the results obtained with the Gaussian pre-training sample, which match the model trained from scratch at small dataset sizes and improve only slightly as more data are added, since the most expressive part of the architecture remains frozen.

Next, we evaluate the predictions of the halo velocities using only the input positions. We employ again the $R^2$ metric to evaluate the results, but sum over all halos and the velocity coordinates, resulting in a single metric value per simulation. The results are shown in Fig.~\ref{fig:camels_velocity}. Similarly to the regression task, we observe good performance from both \cosmos and the model trained from scratch, surpassing previous benchmarks when all available training events are used. This task also highlights the benefits of using more complex deep learning models as opposed to simpler summaries, with the graph neural network (GNN) benchmark introduced in~\cite{Huang:2025dmm} also showing strong performance. The improvements observed from \cosmos compared to the model trained from scratch are reduced, although still present at all dataset sizes. At small dataset sizes, both \cosmos models with full fine-tuning or frozen backbones are significantly better than the model trained from scratch which shows similar performance to the Gaussian pre-training model. This indicates that at small dataset sizes, the learned representation from \cosmos is beneficial to improve the prediction quality whereas the Gaussian pre-training does not yield significant benefits. These results are encouraging and show that models trained on large collider physics datasets encode relevant point cloud structure that can be transferred to other scientific fields.

\begin{figure}[ht]
    \centering
        \includegraphics[width=.45\textwidth]{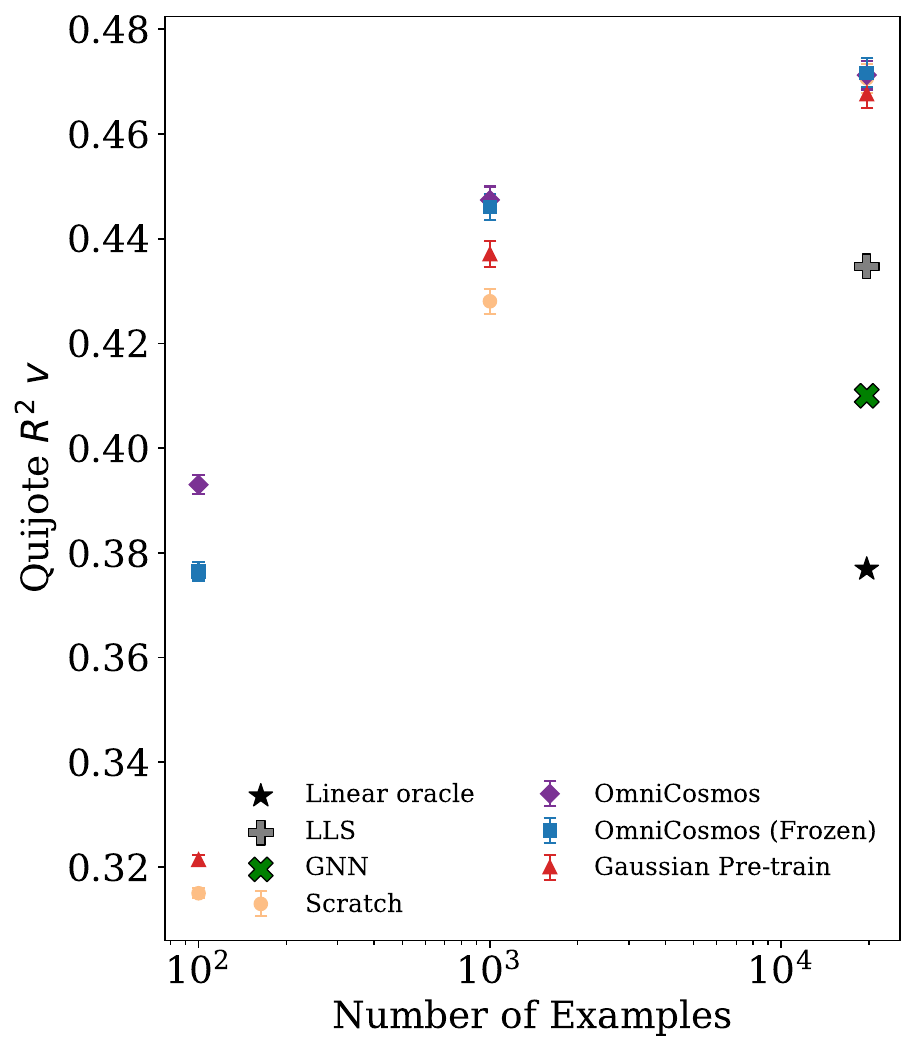}
    \caption{Results for the halo velocity prediction using the \textsc{QUIJOTE} simulations.}
    \label{fig:quijote_velocity}
\end{figure}

Next, we investigate the ability of our model to determine the same set of cosmological parameters but using the \textsc{QUIJOTE} simulations. We increase the number of neighbors considered to $k = 20$, as we observe a small gain in performance using a bigger neighborhood, and scan different hyperparameter settings for both the models training from scratch and different pre-training and adaptation strategies. The results of the best performing models are shown in Table~\ref{tab:quijote_parameter} and Fig.~\ref{fig:quijote_parameter} for the regression of $\Omega_m$ and $\sigma_8$ using different dataset sizes.  Given the bigger simulation samples we can evaluate the model scaling across simulation sizes of different orders of magnitude. Similarly to the results observed using the \textsc{CAMELS-SAM} simulation, the performance obtained by \cosmos either match the best performing benchmarks or surpasses them. In particular, \cosmos is able to match previous benchmarks for $\sigma_8$ using less than $10\%$ of the available simulations. Interestingly, for $\Omega_m$, \cosmos matches the prediction from a two-point correlation function (2PCF), but otherwise greatly improves on the previous deep learning model (GNN). We again observe the model with frozen transformer layers performing equally well as the full fine-tuning \cosmos, in agreement with the results obtained with the \textsc{CAMELS-SAM} simulations. The fine-tuned model pre-trained on the synthetic dataset, however, performs significantly worse. In particular, when only 100 examples are used, the adapted model fails to predict $\sigma_8$ and instead returns approximately the average value present in the training samples, irrespective of the input data. This results in an R$^2$ value close to zero, which is therefore not shown in Fig.~\ref{fig:quijote_parameter}. This behavior remains consistent across all hyperparameter scans performed in this work. These results indicate not only that the synthetic dataset fails to produce useful representations, but also that it harms generalization, leading to worse performance than a model initialized with random weights. Last, we evaluate the velocity prediction using the \textsc{QUIJOTE} simulations with results shown in Fig.~\ref{fig:quijote_velocity}. In previous benchmarks, a simple linear least squares (LLS) model obtained the best predictions for the halo velocities. \cosmos is able to further improve on previous results, surpassing the LLS results using less than $10\%$ of the available dataset. Moreover, we again observe the benefits from the knowledge transfer. Compare to the model trained from scratch, \cosmos is able to show superior performance up to 1000s of simulation samples while converging to the same results when more data are used. Similarly to the \textsc{CAMELS-SAM} results, the model pre-trained on the synthetic Gaussian dataset performs more similarly to the model trained from scratch. In contrast, \cosmos achieves strong performance with both full fine-tuning and frozen transformer layers. The results for the velocity prediction in both datasets studies is listed in Table~\ref{tab:velocities}.

\begin{table}[th]
    \centering
    \caption{Comparison between the performance reported for different  algorithms on the \textsc{QUIJOTE} dataset. Bold results represent the algorithm with highest performance.}
    \label{tab:quijote_parameter}
	\begin{tabular}{lccccc}
    \noalign{\smallskip}\hline
          $\text{R}^2 \uparrow$&  $\Omega_m$ & $\sigma_8$ \\
            \hline
            2PCF & \textbf{0.85 $\pm$ 0.004} & 0.84 $\pm 0.004$\\
            LLS & 0.83 $\pm 0.004$ & 0.80 $\pm 0.004$ \\
            GNN & 0.80 $\pm 0.004$ & 0.77$\pm 0.005$ \\
            GNN (w/o edgeMP) & 0.80 $\pm 0.004$ & 0.79$\pm 0.005$ \\
            \hline
            Scratch & 0.838 $\pm$ 0.003 & 0.868 $\pm$ 0.003   \\
            \textsc{OmniCosmos} & \textbf{0.849 $\pm$ 0.003} & \textbf{0.871 $\pm$ 0.003}   \\
            \textsc{OmniCosmos} (Frozen) & \textbf{0.844 $\pm$ 0.004} & \textbf{0.871 $\pm$ 0.003}   \\
            Gaussian Pre-train & 0.829 $\pm$ 0.004 & 0.855 $\pm$ 0.003   \\
	\noalign{\smallskip}
	\end{tabular}
\end{table}

\begin{table}[th]
    \centering
    \caption{Comparison between the performance reported for different algorithms for halo velocity prediction on the \textsc{CAMELS-SAM} and \textsc{QUIJOTE} datasets. Bold results represent the algorithm with highest performance.}
    \label{tab:velocities}
	\begin{tabular}{lccccc}
    \noalign{\smallskip}\hline
          $\text{R}^2_v \uparrow$&  \textsc{CAMELS-SAM} & \textsc{QUIJOTE} \\
            \hline
            Linear Oracle & 0.2372  & 0.3769\\
            LLS & 0.2107 & 0.4347  \\
            GNN & 0.2865 & 0.4100 \\
            \hline
            Scratch & \textbf{0.298 $\pm$ 0.004} & \textbf{0.470 $\pm$ 0.003 }\\
            \textsc{OmniCosmos} & \textbf{0.298 $\pm$ 0.004 }& \textbf{0.471 $\pm$ 0.003 }\\
            \textsc{OmniCosmos} (Frozen) & \textbf{0.295 $\pm$ 0.004} & \textbf{0.471 $\pm$ 0.003 }   \\
            Gaussian Pre-train & \textbf{0.298 $\pm$ 0.004} & 0.467 $\pm$ 0.003    \\
	\noalign{\smallskip}
	\end{tabular}
\end{table}

\section{Conclusion and Outlook}
\label{sec:conclusions}

In this work, we introduced \cosmos, a machine learning model that leverages a pre-trained network trained on large particle physics datasets to perform different tasks in cosmology. We show how the representation learned by particle interactions can be transferred to a new dataset by quickly adapting the input representation as well as including geometrical insight in the model architecture. Even though the original \omni model was trained on point clouds with one order of magnitude fewer points and for different tasks, the \cosmos model is often able to achieve superior performance compared to the same model trained from scratch. We show the benefits of the pre-training strategy by comparing the results obtained with two additional scenarios: one where the transformer blocks of the model are kept frozen during adaptation, and one where the same architecture is pre-trained on a synthetic Gaussian dataset. We observe that the model with frozen backbone achieves similar performance compared to the full \cosmos model, indicating that the learned relationship between points in the point cloud from collider physics is transferable to cosmology. This observation is further supported by the results obtained using the synthetic dataset, which often under-perform compared to \cosmos. We evaluate \cosmos using simulations available from \textsc{CosmoBench}. In particular, for computationally expensive simulations of galaxy halos such as \textsc{CAMELS-SAM}, \cosmos is able to improve upon all previous benchmarks using half the available dataset. In the \textsc{QUIJOTE} simulations, \cosmos is able to match or surpass all other benchmarks, often using less than $10\%$ of the available data. Since large-scale simulations are computationally expensive, the performance observed by \cosmos has the potential to greatly reduce the number of simulations necessary to achieve precision for cosmology. 

The adaptation step requires fewer than 0.1\% of the original model weights to be replaced with randomly initialized weights. Moreover, while pre-training on collider physics data requires substantial computational resources, adapting the \omni model is considerably less expensive. For instance, adapting \omni to the full \textsc{CAMELS-SAM} dataset takes approximately 8 GPU hours for each task. Although slower, this remains comparable to the 1--3 GPU hours reported for the GNN in~\cite{Huang:2025dmm}. When only 100 simulated samples are used, adaptation takes approximately 4 GPU hours while matching or surpassing all baselines for parameter regression. For the \textsc{QUIJOTE} simulations, adaptation to the full dataset takes approximately 32 GPU hours, comparable to the one-day training time reported in~\cite{Huang:2025dmm}.

Although we observe performance improvements with hundreds of examples, the differences between the model trained from scratch and \cosmos are reduced when thousands of training examples are available. In this work, we explore a simple adaptation strategy where the learning rates of different parts of the model are modified or frozen.  For future work, we plan to investigate alternative strategies to perform the adaptation step, using other techniques introduced in the machine learning community to adapt foundation models to specific problems, such as~\cite{hu2022lora,sun2025transformer}. 

Finally, beyond the regression of cosmological parameters and halo velocity prediction, it would be interesting to directly use the generative capabilities of \omni in the adaptation step of \cosmos for point cloud generation, where the use of a pre-trained model can potentially reduce the number of simulated datasets needed to achieve high fidelity generation, resulting in fast and accurate surrogate models for cosmology.

\section*{Code Availability}

The code for this paper can be found at \url{https://github.com/ViniciusMikuni/OmniLearned}.

\section*{Acknowledgments}
We thank Mariel Pettee for helpful insights during earlier discussions about related projects. We acknowledge the use of Claude 3.5 Sonnet for help building some of the references, which have all been manually checked. VM is supported by JST EXPERT-J, Japan Grant Number JPMJEX2509.
BN is supported by the U.S. Department of Energy (DOE), Office of Science under contract DE-AC02-76SF00515.  This research used resources of the National Energy Research Scientific Computing Center, a DOE Office of Science User Facility supported by the Office of Science of the U.S. Department of Energy under Contract No. DE-AC02-05CH11231 using NERSC awards ERCAP0034229, HEP-ERCAP0021099 and HEP-ERCAP0028249.

\appendix

\bibliography{HEPML,other}
\bibliographystyle{apsrev4-1}

\clearpage

\end{document}